\documentclass[%
aip,
amsmath,amssymb,
reprint,%
]{revtex4-1}
\usepackage{braket}
\usepackage{graphicx}
\usepackage{dcolumn}
\usepackage{bm}

\usepackage[utf8]{inputenc}
\usepackage[T1]{fontenc}
\usepackage{mathptmx}
\usepackage{placeins}  
\usepackage{bibentry}

\usepackage{amsmath}
\usepackage{amsfonts, bbm, accents}

\usepackage[dvipsnames]{xcolor}
\usepackage{svg}
\usepackage[breaklinks=true]{hyperref}
\usepackage{lipsum}
\hypersetup{
  colorlinks   = true, 
  urlcolor     = blue, 
  linkcolor    = blue, 
  citecolor   =  MidnightBlue 
}

\usepackage{soul} 

\usepackage[capitalise]{cleveref} 
\crefformat{equation}{Eq.~(#2#1#3)} 
\crefformat{section}{Sec.~#2#1#3} 
\Crefformat{equation}{Equation~(#2#1#3)}
\crefformat{figure}{Fig.~#2#1#3}
\crefrangeformat{equation}{Eqs.~#3(#1)#4--#5(#2)#6}
\Crefformat{section}{Section~#2#1#3}

\begin{document}


\title{Grain size in low loss superconducting Ta thin films on c-axis sapphire}

\author{Sarah Garcia Jones}
\thanks{These authors contributed equally to this work.}
\affiliation{
Department of Electrical, Computer, and Energy Engineering, University of Colorado Boulder, Boulder, Colorado 80309, USA}%
\affiliation{ 
Department of Physics, University of Colorado, Boulder, Colorado 80309, USA
}%
\author{Nicholas Materise}
\thanks{These authors contributed equally to this work.}
\affiliation{
Department of Physics, Colorado School of Mines, Golden, Colorado 80401, USA}
\author{Ka Wun Leung}
\thanks{These authors contributed equally to this work.}
\affiliation{
Department of Physics, The Hong Kong University of Science and Technology, Clear Water Bay, Kowloon, Hong Kong SAR, China}
\author{Brian D. Isakov}
\affiliation{
Department of Electrical, Computer, and Energy Engineering, University of Colorado Boulder, Boulder, Colorado 80309, USA}
\author{Xi Chen}
\affiliation{
Department of Physics, The Hong Kong University of Science and Technology, Clear Water Bay, Kowloon, Hong Kong SAR, China}
\author{Jiangchang Zheng}
\affiliation{
Department of Physics, The Hong Kong University of Science and Technology, Clear Water Bay, Kowloon, Hong Kong SAR, China}
\author{Andr\'as Gyenis}
\affiliation{
Department of Electrical, Computer, and Energy Engineering, University of Colorado Boulder, Boulder, Colorado 80309, USA}
\author{Berthold Jaeck}
\email[]{bjaeck@ust.hk}
\affiliation{
Department of Physics, The Hong Kong University of Science and Technology, Clear Water Bay, Kowloon, Hong Kong SAR, China}
\affiliation{
HKUST IAS Center for Quantum Technologies, The Hong Kong University of Science and Technology, Clear Water Bay, Kowloon, Hong Kong SAR, China}
\author{Corey Rae H. McRae}
\email[]{coreyrae.mcrae@colorado.edu}
\affiliation{
Department of Electrical, Computer, and Energy Engineering, University of Colorado Boulder, Boulder, Colorado 80309, USA}
\affiliation{ 
Department of Physics, University of Colorado, Boulder, Colorado 80309, USA
}%
\affiliation{ 
National Institute of Standards and Technology, Boulder, Colorado 80305, USA
}%

\date{\today}

\begin{abstract}
In recent years, the implementation of thin-film Ta has led to improved coherence times in superconducting circuits. Efforts to further optimize this materials set have become a focus of the subfield of materials for superconducting quantum computing. It has been previously hypothesized that grain size could be correlated with device performance. In this work, we perform a comparative grain size experiment with $\alpha$-Ta on $c$-axis sapphire. Our evaluation methods include both room-temperature chemical and structural characterization and cryogenic microwave measurements, and we report no statistical difference in device performance between small- and larger-grain-size devices with grain sizes of 924 nm$^2$ and 1700 nm$^2$, respectively. These findings suggest that grain size is not correlated with loss in the parameter regime of interest for Ta grown on c-axis sapphire, narrowing the parameter space for optimization of this materials set.
\end{abstract}


\maketitle 

\section{Introduction}
\label{sec:intro} 
Superconducting qubits are a promising avenue for scalable quantum computing devices due to their high-fidelity operation~\cite{kjaergaard_superconducting_2020,Zhang2021, Hyyppa2022, Howard2022}. Recent advances in qubit design, packaging, and control have shrunken the gap toward their practical use~\cite{preskill2018quantum, arute2019quantum}. Still, dielectric losses due to bulk substrates, surface oxides, and amorphous or defect-ridden material interfaces limit the coherence of superconducting qubits and ancillary devices~\cite{Place2021,Crowley2023,Read2022}. Microscopically, materials loss is largely associated with the excitation of two-level systems (TLS) that dominate microwave losses in the technologically relevant range of low temperatures and single-photon numbers~\cite{Muller2019,McRae2020}. Materials engineering has been identified as a leading route for improvement of superconducting qubit coherence by reducing the effect of TLS~\cite{siddiqi2021engineering}.

Recent works demonstrate improved qubit performance when $\mathrm{\alpha}$-phase tantalum (Ta) replaces niobium (Nb) as the superconducting thin film base layer for device fabrication~\cite{Place2021,Wang2022}. These findings are further supported by loss measurements of superconducting microwave resonators~\cite{Crowley2023,Alegria2023,Shi2022,Lozano2022} and it is believed that the loss reduction is afforded by the simple oxide structure of the Ta film surface~\cite{Place2021}. Further evidence for this is suggested by recent work capping Nb films with Ta for improved qubit performance~\cite{Bal2023}.

A detailed materials study of Nb-based qubits links the bulk properties of the polycrystalline films to qubit losses~\cite{Premkumar2021}. Small crystalline grain sizes were found to correlate with increased qubit losses, which could arise from TLS present at the subsurface grain boundary oxides in Nb films ~\cite{Premkumar2021}. Hence, the grain size of the superconducting base layer has recently been debated as a promising process parameter to further minimize microwave losses in Ta films. Moreover, controlled A/B-testing studies would be desirable to firmly establish this relation and it remains unknown whether grain size effects on microwave losses extend to resonators based on Ta films, whose surface and subsurface oxide structure differs from that of Nb films.

The goal of this work is to probe the relationship between grain size and microwave losses for $\alpha$-Ta films grown on c-axis sapphire, a substrate commonly used for Ta growth~\cite{Place2021,Alegria2023}. To this end, we perform microwave loss measurements of coplanar waveguide resonators made from magnetron-sputtered $\alpha$-Ta films with large and small grain sizes. We compare the losses of both types of films across thirty resonators from multiple chips and report no statistical difference between the performance of films with small and larger grain sizes. In combination with results from the chemical and crystallographic thin film characterizations, our observations indicate that grain size does not play a significant role in microwave losses for $\alpha$-Ta films across the tested parameter regime.
%
\section{Ta Growth and Characterization}
\label{sec:growthandcharacterization} 
%
\begin{figure}[b]
    \centering
    \includegraphics[width=85mm]{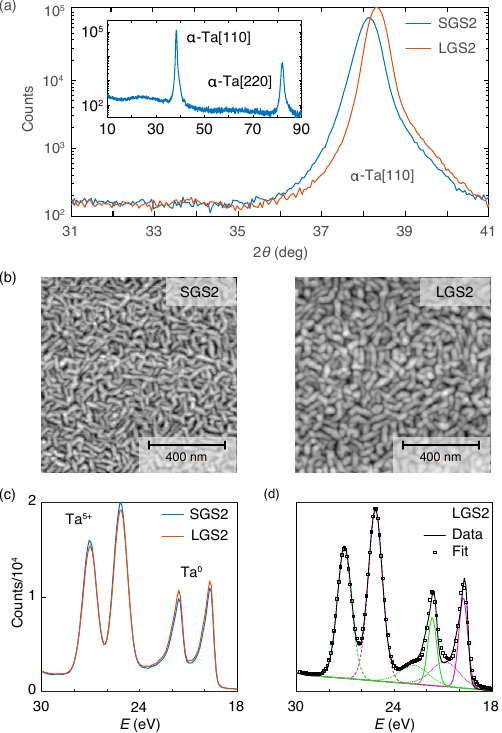}
    \caption{Structural and chemical characterization of Ta films. (a) X-ray diffraction spectra of $2\Theta$-scans for measurements of the `SGS2' and `LGS2' samples. The inset displays the corresponding spectrum for the `LGS2' sample over a larger angle ($\Theta$) range. The detected diffraction peaks are labeled with the corresponding Miller indices of the $\alpha$-Ta phase. (b) Atomic force microscopy topographies of the `SGS2' (left) and `LGS2' (right) sample surfaces. (d) Electron binding energy spectra of the Ta\,4f core level obtained from X-ray photoelectron spectroscopy measurements at the surface of the `SGS2' and `LGS2' samples. The dominant Ta oxidation states are indicated. (d) Least squares fit (open squares) to an XPS spectrum (solid black line) recorded at the surface of the LGS2 sample. Contributions to the spectrum by the Ta\,4f$_{5/2}$ (magenta color) and Ta\,4f$_{7/2}$ (green color) core levels of Ta, (solid lines), Ta$^{3+}$ (dotted lines), and Ta$^{5+}$ (dashed lines) were modeled by using Gaussian profiles.}
\label{fig:structural_chemical_characterization}
\end{figure}

\begin{figure}
    \centering
    \includegraphics[width=85mm]{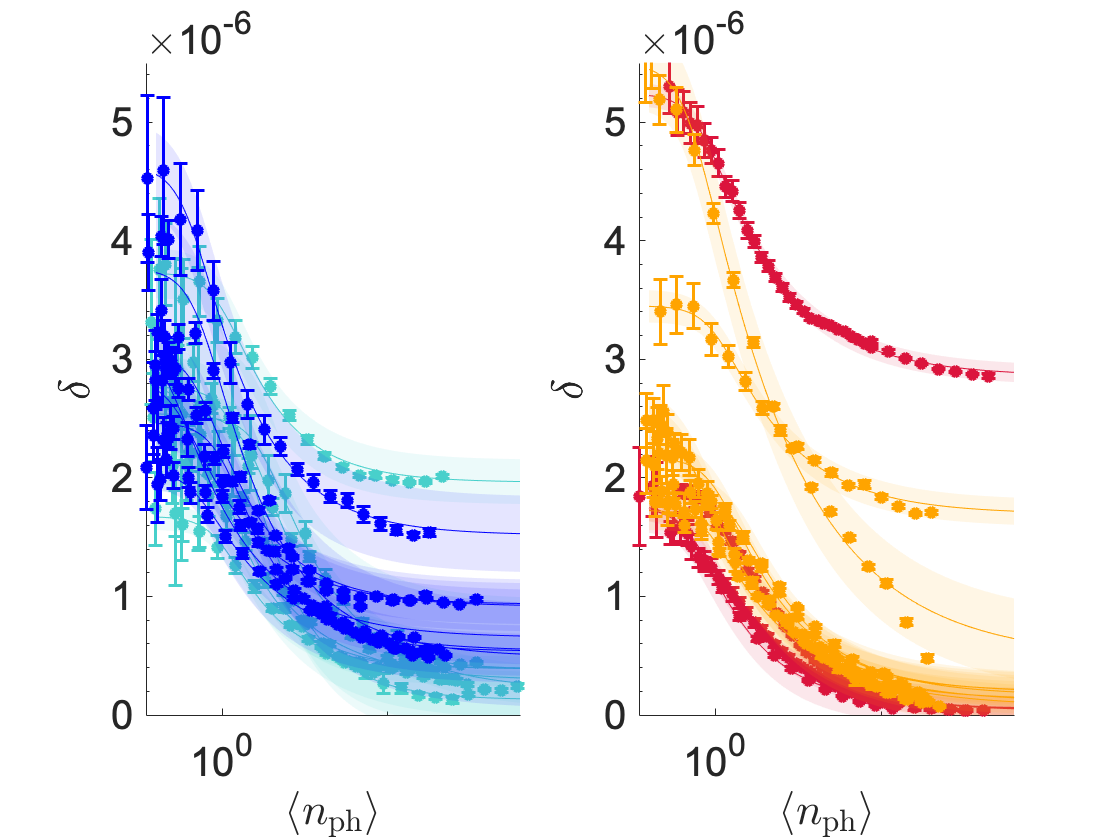}
    \caption{Resonator loss power curves with small grain size (left) and larger grain size (right). Total loss $\delta$ minus high power loss $1/Q_{\mathrm{i,HP}}$ as a function of average number of photons in the cavity $\langle n_{\mathrm{ph}} \rangle$ for all devices measured in this work - SGS1 (light blue), SGS2 (blue), LGS1 (red), and LGS2 (orange)  resonators. Lines denote best fits to the TLS model (Eq.~(\ref{eq:def_delta_tot})). 95$\%$ confidence intervals for Lorentzian fits to each data point are given, as well as the prediction interval for each TLS-curve fit.}
    \label{fig:scurves}
\end{figure}

Ta films of nominal 200 nm thickness were deposited on $c$-axis sapphire wafers (2" diameter, 550\,$\mu$m thickness, from Hefei Keijing Materials Technology) using dc magnetron sputtering. Prior to deposition, the as-purchased substrates were cleaned via ultrasonication in acetone, isopropanol, and deionized water for 5 min each and blown dry with nitrogen of purity 4N. To deposit thin films with different grain sizes, two different substrate temperatures $T=400\,^{\circ}$C (sample label `SGS' or 
`small grain size') and $T=500\,^{\circ}$C (sample label `LGS' or `larger grain size') during the deposition were chosen, while other deposition parameters (background pressure $\leqslant 1 \times 10^{-7}$\,Torr, argon pressure 3\,mTorr, deposition power 150\,W, deposition rate 3.6\,nm/min) were not changed. The deposition was carried out without the use of a seed layer. A $T=600\,^{\circ}$C sample was also grown, but no increase in grain size was detected, so this sample is not included in the detailed film comparison.

The large-scale diffraction spectrum of the films (Fig.\,\ref{fig:structural_chemical_characterization}(a) inset)  is dominated by a set of two peaks, which can be associated with diffraction at the [110] and [220] planes of the $\alpha$-Ta[220] phase. A comparably small diffraction signal, which rises just above the background signal, is detected at 2$\Theta\approx33.7^{\circ}$ that can be associated with the [002]-diffraction of the tantalum $\beta$-phase. Our observations indicate the Ta films prepared for this study predominantly nucleate in the $\alpha$-Ta phase. This finding is consistent with previous reports on 200\,nm thick $\alpha$-Ta films on $c$-axis sapphire, which were deposited under comparable conditions~\cite{Place2021}. The close-up view of the  $\alpha$-Ta[110] peaks for the `SGS2' and `LGS2' samples is shown in the main panel of Fig.~\ref{fig:structural_chemical_characterization}(a). The diffraction peak of the 'LGS2' sample ($\sigma=0.4^{\circ}$) has a smaller full-width-half-maximum $\sigma$ compared to that of the 'SGS2' sample ($\sigma=0.5^{\circ}$). While this observation is indicative of a larger average grain size in the 'LGS2' sample, we note that the Scherrer equation is less suited to quantitatively analyze the grain size in this case, owing to the grain shape anisotropy and significant grain size variations (see AFM measurements below). We further observed a small deviation in the [002]-diffraction angle both between the 'SGS2' (2$\Theta=38.1^{\circ}$) and 'LGS2' (2$\Theta=38.3^{\circ}$) sample, as well as with respect to the nominal bulk value (2$\Theta=38.505^{\circ}$). This can be attributed to the presence of strain in the thin film structure, which appears slightly more pronounced in the `SGS2' sample.

To characterize the crystalline grain size of the Ta films deposited at different substrate temperatures, we carried out atomic force microscopy (AFM) measurements (tapping mode). The resulting AFM topographies for samples `SGS2' and `LGS2' are shown in Fig.\,\ref{fig:structural_chemical_characterization}(b). Both topographies are characterized by elongated crystalline grains oriented along the hexagonal basal plane of the sapphire surface, consistent with previous reports~\cite{Alegria2023}. Moreover, the grains of `LGS2' exhibit a visibly larger grain size area $G$ than those of `SGS2', consistent with our expectations in light of the substrate temperatures during deposition. To quantify these grain size differences, we applied a watershed algorithm~\cite{rabbani2015} to determine $G$, which is an average across several 1\,$\mu$m$^2$ surface areas per sample and several samples for each deposition condition. This approach was previously applied to quantify grain sizes of Nb films~\cite{Premkumar2021}. We obtain $G=924\pm51 \,\mathrm{nm}^2$ for the `SGS2' and $G=1700\pm29\,\mathrm{nm}^2$ for the `LGS2' sample, respectively. Interestingly, the average grain size $G=1732\pm92\,\mathrm{nm}^2$ of samples deposited at a substrate temperature $T=600\,^{\circ}$C is comparable to that of the $T=500\,^{\circ}$C deposition~\cite{SI}.

To detect the possible influence of the crystalline grain size on the surface oxide structure, we performed X-ray photoelectron spectroscopy (XPS) measurements (Kraxios Ultra DLD; X-ray source: Al K$\alpha$ line $E=1486.6\,$eV) on the `SGS2' and `LGS2' samples. We note that these samples did not undergo surface treatment to remove native surface oxides prior to XPS measurements. The resulting XPS spectra in Fig.\,\ref{fig:structural_chemical_characterization}(c) show the photo-electron count as a function of the electron binding energy for the Ta-4f core level. The spectra are dominated by a four peak structure, which is predominantly composed of the spin-orbit split Ta$^0$ and Ta$^{5+}$ doublets that can be assigned to the metallic Ta bulk and the Ta$_2$O$_5$ at the film surface, respectively~\cite{mcguire1973core, himpsel1984core}. 

We quantify the relative contributions of the different Ta oxidation states to the observed XPS spectra by applying a least-squares fit based on Gaussian profiles. We find a three doublet structure composed of six Gaussians, as shown in Fig.\,\ref{fig:structural_chemical_characterization}(d), can most accurately describe these spectra. The additional third doublet exhibits a core level shift of $\approx1.1\,$eV and can be assigned to the Ta$^{3+}$ oxidation state of the Ta$_2$O$_3$ suboxide~\cite{himpsel1984core}. The resulting relative contributions of Ta, Ta$^{3+}$, and Ta$^{5+}$ obtained from these fits are shown in Table\,\ref{tab:xps} and reveal a near identical chemical structure of the tantalum film surface for both samples. this is consistent with their almost identical XPS spectra ({\em cf.}~Fig.~\ref{fig:structural_chemical_characterization}(c)). The relative spectral weight of the Ta$^0$ and Ta$^{5+}$ peaks at the given incident X-ray energy is in close agreement with that found in previous XPS studies of tantalum films and indicates a surface oxide thickness of approximately 2\,nm~\cite{Place2021}.

\begin{table}
    \caption{\label{tab:xps}Relative atomic concentration of different tantalum oxidation states in the 'SGS2' and 'LGS2' samples as obtained from fits to the XPS spectra.}
    \begin{ruledtabular}
        \begin{tabular}{lcc}
         & Small grain size & Larger grain size\\
        \hline
        Oxidation state & atomic \% & atomic \% \\
        \hline
        Ta$^0$ & 18 & 20\\
        Ta$^{3+}$  & 17 & 18 \\
        Ta$^{5+}$ & 65 & 62 \\
        \end{tabular}
    \end{ruledtabular}
\end{table}

%
\section{Device Design and Fabrication}
\label{sec:devicefab} 
\begin{figure}[!ht]
    \centering
    \includegraphics[width=85mm]{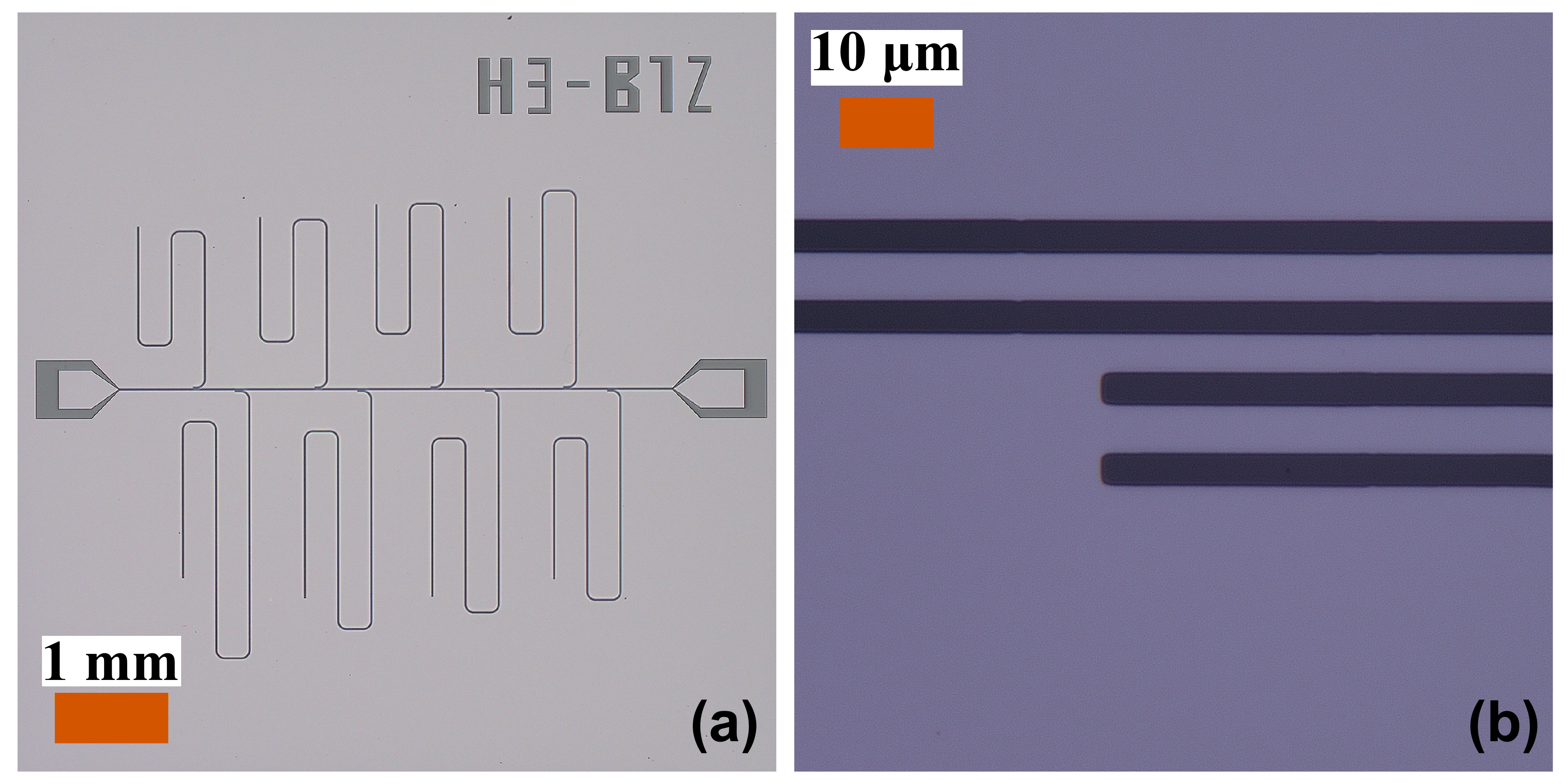}
    \caption{Optical microscope images of coplanar waveguide resonators. (a) Full chip image of a representative chip. All circuits measured contain eight resonators with identical couplers. (b) Close up of feedline and resonator base. Conductor width is 5.5 $\mu m$ and gap is 3.8 $\mu m$.}
\label{fig:3Z_optical}
\end{figure}

All devices are coplanar waveguide resonators fabricated using the same designs as reported by~\citeauthor{Kopas2022}~\cite{Kopas2022}. Nominally identical designs and fabrication procedures were used for all samples.  Prior to etching, the samples were cleaned via ultrasonication in toluene, acetone, methanol, and isopropanol, then patterned using optical lithography and AZ-P4330-RS photoresist. The films were etched in a single 4 minute CF$_{4}$/N$_{2}$ Inductively Coupled Plasma – Reactive Ion Etch (Panasonic E640).  Since the Ta films were deposited on sapphire substrates, the etches did not produce any trenching into the substrate. After etching, the resist was submerged in AZ 300 T stripper at 80 $^{\circ}$C for 1 hour.  After stripping, the samples were diced and again cleaned ultrasonically in toluene, acetone, methanol, and isopropanol before being wire bonded for cryogenic microwave measurement. Optical images of the resonators are shown in Fig.~\ref{fig:3Z_optical}.

Inverse coupling quality factors, $1/Q_c$, of the fabricated resonators are presented in Supplementary Materials Table 1 and range from $1.18 \times 10^{-6}$ to $6.61 \times 10^{-6}$ across all devices. This is a larger spread of values with a trend towards smaller coupling factors than the simulated $1/Q_c$ values of these designs, which ranged from $1.95 \times 10^{-6}$ to $2.02 \times 10^{-6}$.~\cite{Kopas2022} This variation is likely due to a slight over etch of the devices during fabrication, which is congruent with a thinner measured conductor width than the lithography designs used (design: $6 \mu\, \mathrm{m}$, measured: $5.5 \mu\,\mathrm{m}$).

%

\section{Cryogenic Microwave Measurement}
\label{sec:cryogenicmicrowavemeasurement} 

\begin{figure*}[t]
    \centering
    \includegraphics[width=160mm]{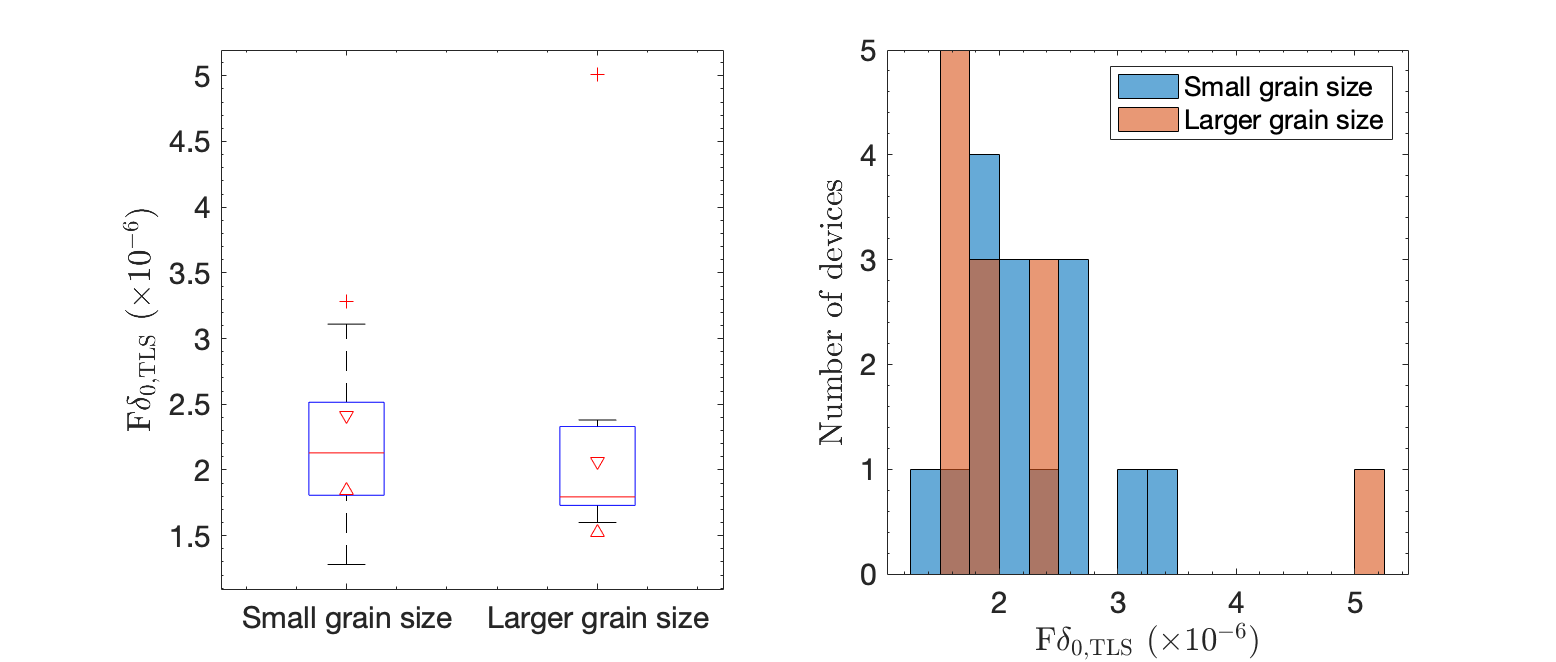}
    \caption{TLS loss in Ta on c-axis sapphire superconducting microwave resonators with small and larger grain size. Left: Box and whiskers comparison, indicating median values (red line) and 95$\%$ confidence interval of median (red triangles), with outliers shown as crosses. Right: Histogram of TLS loss for all devices in this experiment.}
    \label{fig:boxwhiskers}
\end{figure*}
We perform transmission measurements on CPW resonators mounted to the mixing chamber (MXC) plate of a FormFactor (formerly Janis) JDry 250 dilution refrigerator (DR) at a mixing chamber temperature of $\sim$10 mK using a Keysight PNA N5222B vector network analyzer (VNA). The input power varied over ten orders of magnitude in estimated photon power to accurately extract the dominant two level system (TLS) loss~\cite{GaoThesis}. 

Nominally identical gold-plated oxygen-free high conductivity copper sample boxes house the resonator chips and are mounted to a plate perpendicular to the MXC plate with additional mu-metal shielding surrounding all samples. Two Radiall R583 six-way microwave switches allow multiple samples to be measured on the same pair of coaxial input and output lines.

The transmission data for each resonator ($S_{21}$ of the two-port $S$-parameter matrix measured by the VNA) is first normalized with a circle fit~\cite{Probst2015}, and then fit to a diameter-corrected asymmetric Lorentzian model of the form~\cite{Khalil2012},
    \begin{eqnarray}
       S_{21}(f) &=& 1 - \frac{Q_l\ /\ Q_c\ e^{i\phi}}{1 + 2iQ \frac{f - f_0}{f_0}}\label{eq:S21_DCM} \\
       Q_i^{-1} &=& Q_l^{-1} - \mathrm{Re}\left\{\hat{Q}^{-1}_c\right\} \label{eq:Ql_Qi_Qc} \\
       \hat{Q}_c^{-1} &=& Q_c^{-1} e^{i\phi} \label{eq:def_Qchat}
    \end{eqnarray}
where $f_0$ is the resonance frequency, $\phi$ is the asymmetry angle, $Q_c$ is coupling quality factor, $Q_l$ is the loaded quality factor, and $Q_i$ is the internal quality factor. These parameters are fit with their corresponding 95\% confidence intervals from a least squares fitting routine~\cite{BCQT_Code}. A secondary fit of the loss $\delta = Q_i^{-1}$ as a function of average number of photons $\left<n_{\mathrm{ph}}\right>$ and fixed temperature $T$ follows from the sum of the TLS loss contribution $\delta_{\mathrm{TLS}}$ and an offset term $1/Q_{\mathrm{HP}}$ that accounts for power-independent losses dominating at higher powers~\cite{McRae2020}
    \begin{eqnarray}
        \delta\left(\left<n_{\mathrm{ph}}\right>, T\right) &=&
        \delta_{\mathrm{TLS}}\left(\left<n_{\mathrm{ph}}\right>, T\right) + 1/Q_{\mathrm{HP}}
        \label{eq:def_delta_tot} \\
        \delta_{\mathrm{TLS}}\left(\left<n_{\mathrm{ph}}\right>, T\right)
        &=& F \delta_{\mathrm{TLS}}^0\frac{\tanh\left(\frac{\hbar\omega_0}{k_B T}\right)}%
        {\left(1 + \frac{\left<n_{\mathrm{ph}}\right>}{n_c}\right)^{\beta}}
        \label{eq:def_delta_tls}
    \end{eqnarray}
where $n_c$ is the critical photon number at which TLS saturate at low power, $\omega_0=2\pi f_0$ is the angular resonance frequency, $\beta$ is an exponent interpolating between the non-interacting TLS model $\beta=1/2$ and interacting TLS model $\beta < 1/2$~\cite{GaoThesis,Phillips1972, DeGraaf2018}, $\delta_{\mathrm{TLS}}^0$ is the intrinsic TLS loss, $F$ is the geometry-dependent filling factor, $\hbar$ is the reduced Planck constant, and $k_B$ is the Boltzmann constant.

In Fig.~\ref{fig:scurves}, we plot the loss power dependence for all large and small grain size devices used in this work. The high power losses are subtracted to emphasize the similar power dependence (line shape) and low power loss (TLS saturation loss) between the two samples, without the confounding factor of high-power losses which are known to be caused by a myriad of sources external to the device materials~\cite{McRae2020}. Figure~\ref{fig:boxwhiskers} further highlights this point, as the medians from the box and whisker plots of the intrinsic TLS losses for the large and small grain size Ta films coincide with one another and their respective histograms give similar variances. 
\begin{table}
    \caption{\label{tab:ressum}Mean parameter values in A/B grain size comparison.}
    \begin{ruledtabular}
        \begin{tabular}{ccc}
         & Small grain size & Large grain size\\
        \hline
        Grain area ($\mathrm{nm^2}$) & $924\pm51$ & $1700\pm29$\\
        $F \delta_{\mathrm{TLS}}^{0}$ ($\times 10^{-6}$) & 2.19 $\pm$ 0.07 & 2.17 $\pm$ 0.03 \\
        \end{tabular}
    \end{ruledtabular}
\end{table}
%
\section{Literature Comparison}
\label{sec:statisticalanalysis} 

\begin{figure}[b]
    \centering
    \includegraphics[width=90mm]{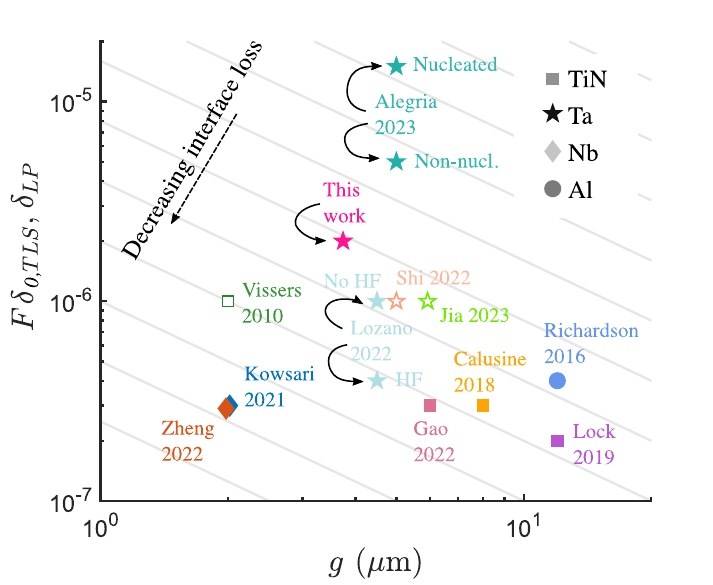}
    \caption{State-of-the-art literature comparison of TLS ($F \delta_{\mathrm{TLS}}^{0}$) and low-power ($\delta_{\mathrm{LP}}$) loss values in CPW resonators. Filling factor $F$ is estimated by plotting loss as a function of CPW gap width $g$. Grey lines denote lines of constant interface loss. Filled symbols denote TLS loss values, while empty symbols represent low power loss (TLS loss values unavailable).}
    \label{fig:litscatterplot}
\end{figure}

Mean parameter values for the two resonator populations are summarized in Tab.~\ref{tab:ressum}. Though our small-geometry resonators are very sensitive to TLS loss, we see no statistical difference in TLS loss between the small-grain and large-grain devices, despite the difference in grain size (with the large grain size devices being almost twice as big in area). This difference in grain size is similar to that seen in Nb films in Ref.~\cite{Premkumar2021}, where a difference in grain size of a factor of two was correlated with a difference in qubit $T_1$ of almost two, and a difference in resonator TLS loss was also detected.

We refer to recent studies, especially the work by \citet{Lozano2022}, to estimate the number of devices required to adequately sample the device-to-device variation in $F\delta_{\mathrm{TLS}}^0$. With more than ten devices of each grain size, we exceed the number of devices for each variation (choice of control parameter value) in Refs.~\citet{Lozano2022, Alegria2023, Crowley2023, Jia2023}.

The median values of the losses for the untreated Ta resonators on silicon in Ref.~\citet{Lozano2022} are comparable to the large- and small-grain size median losses reported in Fig.~\ref{fig:boxwhiskers}(a). 

Figure.~\ref{fig:litscatterplot} shows that the difference between the large- and small-grain size losses in this study is imperceptible on the same scale as other A/B comparisons, e.g. Refs.~\citet{Alegria2023, Lozano2022}
%
\section{Discussion}
\label{sec:discussion} 
An enduring hypothesis in the superconducting qubit community has been that larger grain size in superconducting thin films is an indication of improved device performance. The simple and stable oxide structure of Ta differs from that of Nb, where~\citet{Premkumar2021} reported that smaller grain size films exhibited higher concentrations of suboxides in interface regions, resulting in measurably higher losses in their Nb resonators. In this study, we show that smaller grain size does not induce significant low power loss in Ta thin films grown between 400 and 500\,°C on c-axis sapphire. 

Microwave measurements of low power loss suggest that there is no statistically significant difference between the intrinsic TLS losses of the two grain size Ta thin films. Chemical and structural analysis support this interpretation, as the surface chemistry obtained by XPS is nearly identical for the two films. This distinguishes densely packed Ta films with their simple Ta$_2$O$_5$ surface oxide structure~\cite{Place2021} from Nb films~\cite{Premkumar2021} for which subsurface grain boundary oxides contribute a grain size dependent TLS channel. Following this train of thought, we expect qubits and resonators fabricated from Ta films to exhibit more uniform microwave losses than those fabricated from Nb. At the same time, non-negligible concentrations of Ta$^{3+}$ species found in our XPS measurements indicates the presence of Ta$_2$O$_3$ suboxides at the Ta metal-Ta$_2$O$_5$ interface consistent with a recent report~\cite{mclellan2023}. Interestingly, we also detect practical limits within which to tune the grain size of [110]-oriented $\alpha$-Ta films deposited on $c$-axis sapphire: Deposition at substrate temperatures below 400\,°C favors the formation of the unwanted $\beta$-phase~\cite{knepper2006effect, myers2013beta} whereas grain size does not respond to an increase of substrate temperature in excess of 500\,°C in our study. Thus, it would be interesting to explore other substrates or sapphire surface orientations to promote larger grain sizes up to the formation of single-crystalline Ta films. On the other hand, our study suggests more sophisticated materials engineering efforts that focus on the reduction of TLS losses at the immediate metal-air surface rather than on the optimization of bulk properties, such as grain size, are required to further reduce microwave losses below those reported in this and other recent studies~\cite{Lozano2022, Alegria2023}. These efforts will benefit from targeted A/B testing studies, such as is presented here, to address the vast materials and processing parameter space in order to maximize state-of-the-art superconducting qubit performance.
%
\section{Conclusion}
\label{sec:conc} 
We performed millikelvin microwave transmission measurements of $\alpha$-phase Ta microwave resonators with both large- and small-grain size sputter-deposited on c-axis sapphire at two different growth temperatures. Structural and chemical analysis reveal that the films differ only in their grain size and not in their surface oxide types and concentrations, and crystal structure. The extracted intrinsic TLS losses show no statistical difference between the two film types, suggesting that, in this materials regime, grain size does not significantly affect millikelvin, ultralow power dieletric loss. We encourage future A/B experimentation to continue to reduce the fabrication parameter space and to identify correlations between other room-temperature materials characterization parameters and low-power, low-temperature microwave performance of devices.
%
\begin{acknowledgments}
The authors would like to thank Qiming Shao and Jiacheng Liu for their assistance with the sputter deposition, and Carlos Torres Castanedo, David Garcia, Tony McFadden, and Dominic Goronzy for helpful discussions and feedback on the manuscript. N.M. acknowledges funding from the Graduate Fellowships for STEM Diversity. B.J. acknowledges funding from the Croucher Foundation.
\end{acknowledgments}
%
\section*{Author Declarations}

\subsection*{Conflicts of Interest}
The authors have no conflicts of interest.

\subsection*{Commercial Products Disclaimer}

Certain commercial instruments are identified to specify the experimental study adequately. This does not imply endorsement by NIST or that the instruments are the best available for the purpose.
%
\subsection*{Author Contributions}
\textbf{Sarah Garcia Jones}: Resonator sample fabrication (lead), Writing - review \& editing; \textbf{Nicholas Materise}: Writing - original draft (lead), Writing - review \& editing (supporting), Data acquisition (equal), Software (equal); \textbf{Ka Wun Leung}: Materials fabrication (lead) ; \textbf{Brian D. Isakov}: Writing - review \& editing; \textbf{Chen Xi}: Materials characterization (lead); \textbf{Andr\'{a}s Gyenis}: Writing - review \& editing, Resonator sample fabrication; \textbf{Bertold Jaeck}: Conceptualization (equal), Writing - review \& editing (lead), Visualization (equal), Formal analysis (equal), Methodology (lead); \textbf{Corey Rae H. McRae}: Conceptualization (equal), Writing - review \& editing (lead), Visualization (equal), Formal analysis (equal), Methodology (lead).

\section*{Data Availability}
The data that support the findings of this study are openly available at https://zenodo.org/record/8161535, reference number 10.5281/zenodo.8161535.

\section*{References}
\bibliography{TaGrowth2023.bib}

\end{document}



\title{Supplementary Material: Grain size in low loss superconducting Ta thin films on c-axis sapphire}

\author{Sarah Jones*}
\affiliation{
Department of Electrical, Computer, and Energy Engineering, University of Colorado Boulder, Boulder, Colorado 80309, USA}
\author{Nicholas Materise*}
\affiliation{
Department of Physics, Colorado School of Mines, Golden, Colorado 80401, USA}
\author{Ka Wun Leung*}
\affiliation{
Department of Physics, The Hong Kong University of Science and Technology, Clear Water Bay, Kowloon, Hong Kong SAR, China}
\author{Brian Isakov}
\affiliation{
Department of Electrical, Computer, and Energy Engineering, University of Colorado Boulder, Boulder, Colorado 80309, USA}
\author{Xi Chen}
\affiliation{
Department of Physics, The Hong Kong University of Science and Technology, Clear Water Bay, Kowloon, Hong Kong SAR, China}
\author{Andr\'as Gyenis}
\affiliation{
Department of Electrical, Computer, and Energy Engineering, University of Colorado Boulder, Boulder, Colorado 80309, USA}
\author{Berthold Jaeck}
\email[]{bjaeck@ust.hk}
\affiliation{
Department of Physics, The Hong Kong University of Science and Technology, Clear Water Bay, Kowloon, Hong Kong SAR, China}
\affiliation{
HKUST IAS Center for Quantum Technologies, The Hong Kong University of Science and Technology, Clear Water Bay, Kowloon, Hong Kong SAR, China}
\author{Corey Rae H. McRae}
\email[]{coreyrae.mcrae@colorado.edu}
\affiliation{
Department of Electrical, Computer, and Energy Engineering, University of Colorado, Boulder, Colorado 80309, USA}
\affiliation{ 
National Institute of Standards and Technology, Boulder, Colorado 80305, USA
}%
\affiliation{ 
Department of Physics, University of Colorado, Boulder, Colorado 80309, USA
}%

\date{\today}

\pacs{}

\maketitle 


\section{Ta grain size for T=600\,°C deposition}
\begin{figure}[b]
    \centering
    \includegraphics[width=80mm]{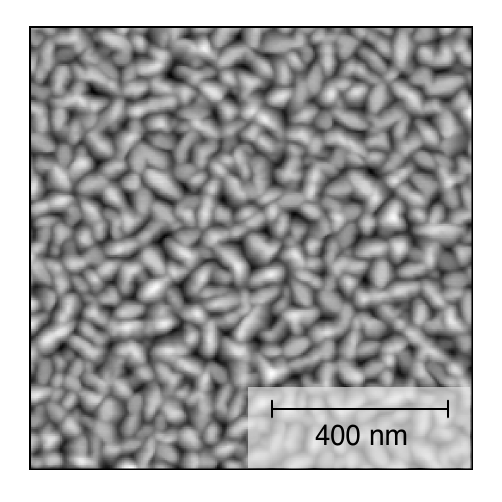}
    \caption{Typical surface topography recorded with an atomic force microscope of the tantalum film deposited at a substrate temperature of T=600\,°C.}
    \label{fig:600degGS}
\end{figure}

We used atomic force microscopy in tapping mode to study the grain size distribution of the tantalum films deposited at a substrate temperature of T=600\,°C. A typical topography is shown in Fig.~\ref{fig:600degGS}, with an average grain size of $G=1732\pm92\,\mathrm{nm}^2$.

\section{Wide scans}
\begin{figure*}[t]
    \centering
    \includegraphics[width=170mm]{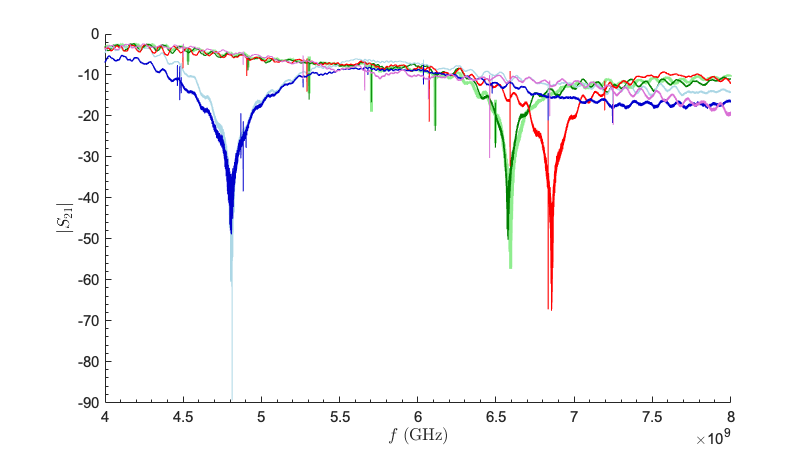}
    \caption{Background characterization at -100 dBm. Transmission curves for SGS1 cooldown 2 (light blue) and 4 (blue), LGS1 cooldown 1 (light green) and 3 (green), LGS2 cooldown 4 (purple) and SGS2 cooldown 4 (red) are plotted.}
    \label{fig:widescan}
\end{figure*}

We measured the broadband responses of the devices over multiple cooldowns to compare the variability of the background (Fig.~\ref{fig:widescan}). We measured $S_{21}$, the microwave transmission coefficient, through each device's transmission line with 65,001 points and 100 kHz IF bandwidth to resolve most of the resonator dips over the 4--8 GHz range. Repeated measurements of the same devices show significant similarities, as shown by the two blue curves (SG2) and the two green curves (LGS1). Broad resonances can be attributed to a low-Q spurious environmental mode, which is present for all but LGS2 (purple curve).

\section{Extracted resonator parameters}
Table~\ref{tab:resmeas} summarizes the parameters extracted from fitting the TLS and power-independent loss of each resonator. The uncertainties are the 95\% confidence intervals returned by a least squares fitting routine. 

\begin{figure*}[t]
    \centering
    \includegraphics[width=170mm]{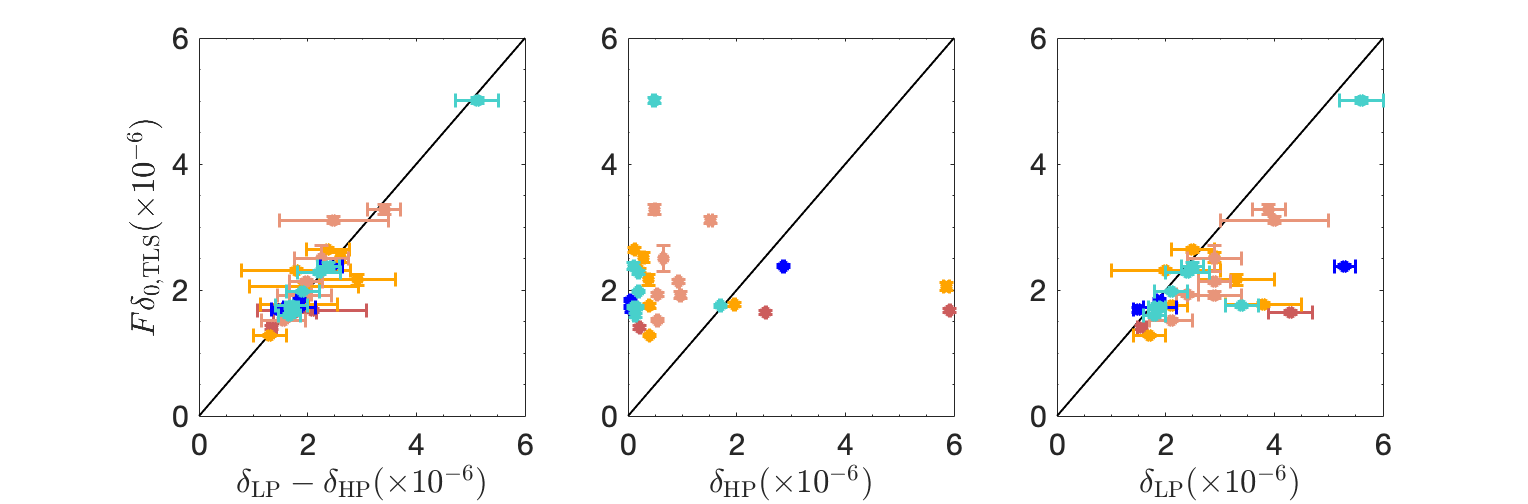}
    \caption{Correlation of loss metrics. Device-induced intrinsic TLS loss $F \delta_{\mathrm{0,TLS}}$ as a function of three other loss metrics: low power loss $\delta_{\mathrm{LP}}$ minus power-independent loss $\delta_{\mathrm{HP}}$, $\delta_{\mathrm{HP}}$, and $\delta_{\mathrm{LP}}$. Solid black line indicates a 1:1 relationship between metrics.}
    \label{fig:LPminusHP}
\end{figure*}

Figure~\ref{fig:LPminusHP} compares the loss metrics $F \delta_{\mathrm{TLS}}^0$ with $\delta_{\mathrm{LP}}$, $\delta_{\mathrm{HP}}$, and $\delta_{\mathrm{LP}}-\delta_{\mathrm{HP}}$. There is excellent agreement between $F \delta_{\mathrm{TLS}}^0$ and $\delta_{\mathrm{LP}}-\delta_{\mathrm{HP}}$, where all measurements lay along the line of 1:1 correlation. The mismatch between $F \delta_{\mathrm{TLS}}^0$ and $\delta_{\mathrm{LP}}$ highlights that one cannot extract the intrinsic TLS loss with the low power loss alone; the power-independent loss is necessary to accurately compute TLS loss. The scatter in the center plot indicates a lack of correlation between power-independent loss and TLS loss.

\begin{table*}[t]
\caption{\label{tab:resmeas}Parameters extracted from cryogenic microwave measurements of Ta on Al2O3 coplanar waveguide (CPW) resonators. Values are given with their 95$\%$ confidence intervals where available. $f_0$: resonance frequency. $1/Q_{i,\mathrm{HP}}$: inverse high power internal quality factor. $F \delta_{\mathrm{TLS}}^{0}$: resonator-induced intrinsic TLS loss. $1/Q_{i,\mathrm{LP}}$: inverse low power internal quality factor. $1/Q_c$: inverse coupling quality factor at high power. Surface treatment labels correspond to small grain size (SGS*) and large grain size (LGS*) devices.}
\begin{ruledtabular}
\begin{tabular}{ccccccccc}
Surface treatment & $f_0$ (GHz) & $1/Q_{i,\mathrm{HP}}$ ($\times 10^{-6}$) & $F \delta_{\mathrm{TLS}}^{0}$ ($\times 10^{-6}$) & $1/Q_{i,\mathrm{LP}}$ ($\times 10^{-6}$) & $1/Q_c$ ($\times 10^{-6}$) & $n_c$ & $\beta$\\
\hline
LGS1 & 5.704 & 0.05 $\pm$ 0.01 & 1.73 $\pm$ 0.02 & 1.9 $\pm$ 0.3 & 4.658 $\pm$ 0.004 & 0.14 $\pm$ 0.02 & 0.197 $\pm$ 0.005 \\
Cooldown 1 & 6.112 & 0.057 $\pm$ 0.005 & 1.69 $\pm$ 0.04 & 1.5 $\pm$ 0.1 & 5.702 $\pm$ 0.002 & 0.19 $\pm$ 0.06 & 0.24 $\pm$ 0.02 \\
\hline
LGS1 & 4.531  & 0.037 $\pm$ 0.008 & 1.84 $\pm$ 0.02 & 1.89 $\pm$ 0.06 & 1.871 $\pm$ 0.004 & 1.8 $\pm$ 0.2 & 0.237 $\pm$ 0.005 \\
Cooldown 3 & 4.915 & 2.86 $\pm$ 0.02 & 2.38 $\pm$ 0.02 & 5.3 $\pm$ 0.2 & 2.649 $\pm$ 0.009 & 0.34 $\pm$ 0.03 & 0.195 $\pm$ 0.002 \\
\hline
LGS2 & 4.501 & 0.10 $\pm$ 0.02 & 1.73 $\pm$ 0.03 & 1.79 $\pm$ 0.1 & 1.199 $\pm$ 0.004 & 0.4 $\pm$ 0.1 & 0.175 $\pm$ 0.009 \\
Cooldown 4 & 4.884 & 0.48 $\pm$ 0.03 & 5.01 $\pm$ 0.05 & 5.6 $\pm$ 0.4 & 1.46 $\pm$ 0.01 & 0.17 $\pm$ 0.04 & 0.151 $\pm$ 0.006 \\
 & 5.267 & 0.110 $\pm$ 0.009 & 2.38 $\pm$ 0.05 & 2.5 $\pm$ 0.2 & 1.892 $\pm$ 0.004 & 0.10 $\pm$ 0.04 & 0.18 $\pm$ 0.01 \\
 & 5.661 & 0.19 $\pm$ 0.02 & 1.98 $\pm$ 0.02 & 2.1 $\pm$ 0.3 & 3.28 $\pm$ 0.01 & 1.0 $\pm$ 0.2 & 0.212 $\pm$ 0.008 \\
 & 6.068 & 0.17 $\pm$ 0.01 & 1.73 $\pm$ 0.02 & 1.9 $\pm$ 0.1 & 2.088 $\pm$ 0.005 & 3.6 $\pm$ 0.8 & 0.21 $\pm$ 0.01 \\
 & 6.459 & 0.13 $\pm$ 0.02 & 1.60 $\pm$ 0.03 & 1.8 $\pm$ 0.2 & 6.14 $\pm$ 0.03 & 1.4 $\pm$ 0.4 & 0.21 $\pm$ 0.01 \\
 & 6.844 & 1.70 $\pm$ 0.02 & 1.75 $\pm$ 0.02 & 3.4 $\pm$ 0.3 & 5.71 $\pm$ 0.02 & 1.0 $\pm$ 0.2 & 0.207 $\pm$ 0.008 \\
 & 7.250 & 0.19 $\pm$ 0.02 & 2.28 $\pm$ 0.03 & 2.4 $\pm$ 0.4 & 3.76 $\pm$ 0.01 & 0.13 $\pm$ 0.03 & 0.201 $\pm$ 0.008 \\
\hline\hline
SGS1 & 4.488 & 0.22 $\pm$ 0.01 & 1.41 $\pm$ 0.03 & 1.56 $\pm$ 0.08 & 1.184 $\pm$ 0.004 & 1.5 $\pm$ 0.3 & 0.236 $\pm$ 0.008 \\
Cooldown 2 & 5.682 & 5.92 $\pm$ 0.01 & 1.68 $\pm$ 0.03 & 8 $\pm$ 1 & 1.631 $\pm$ 0.002 & 0.23 $\pm$ 0.05 & 0.24 $\pm$ 0.01 \\
& 6.476 & 2.535 $\pm$ 0.006 & 1.64 $\pm$ 0.03 & 4.3 $\pm$  0.4 & 1.663 $\pm$ 0.002 & 0.32 $\pm$ 0.07 & 0.26 $\pm$ 0.01 \\
\hline
SGS1 & 4.487$^{\ast}$ & 0.38 $\pm$ 0.02 & 2.16 $\pm$ 0.09 & 3.3 $\pm$ 0.7 & 1.244 $\pm$ 0.008 & 9 $\pm$ 7 & 0.27 $\pm$ 0.06  \\
Cooldown 4 & 4.902 & 0.21 $\pm$ 0.01 & 2.31 $\pm$ 0.03 & 2 $\pm$ 1 & 1.85 $\pm$ 0.01 & 15 $\pm$ 3 & 0.20 $\pm$ 0.01 \\
& 5.267 & 0.128 $\pm$ 0.010 & 2.64 $\pm$ 0.04 & 2.5 $\pm$ 0.4 & 1.68 $\pm$ 0.01 & 3.4 $\pm$ 0.9 & 0.29 $\pm$ 0.02 \\
& 5.681$^{\ast}$ & 5.86 $\pm$ 0.06 & 2.06 $\pm$ 0.06 & 7.8 $\pm$ 1.0 & 1.576 $\pm$ 0.004 & 0.5 $\pm$ 0.2 & 0.25 $\pm$ 0.02  \\
& 6.037 & 0.395 $\pm$ 0.007 & 1.28 $\pm$ 0.02 & 1.7 $\pm$ 0.3 & 1.659 $\pm$ 0.004 & 0.9 $\pm$ 0.2 & 0.28 $\pm$ 0.02 \\
& 6.476$^{\ast}$ & 1.96 $\pm$ 0.01 & 1.77 $\pm$ 0.03 & 3.8 $\pm$ 0.7 & 1.522 $\pm$ 0.006 & 0.9 $\pm$ 0.3 & 0.27 $\pm$ 0.02  \\
& 6.833 & 0.29 $\pm$ 0.02 & 2.52 $\pm$ 0.08 & 2.9 $\pm$ 0.04 & 1.99 $\pm$ 0.02 & 0.13 $\pm$ 0.06 & 0.21 $\pm$ 0.02 \\
& 7.245 & 0.39 $\pm$ 0.01 & 1.75 $\pm$ 0.04 & 2.1 $\pm$ 0.3 & 1.269 $\pm$ 0.008 & 0.9 $\pm$ 0.2 & 0.29 $\pm$ 0.02 \\
\hline
SGS2 & 4.499 & 0.65 $\pm$ 0.01 & 2.5 $\pm$ 0.2 & 2.9 $\pm$ 0.5 & 2.106 $\pm$ 0.006 & 0.020 $\pm$ 0.018 & 0.21 $\pm$ 0.02 \\
Cooldown 4 & 4.907 & 0.49 $\pm$ 0.01 & 3.28 $\pm$ 0.08 & 3.9 $\pm$ 0.3 & 2.502 $\pm$ 0.007 & 0.210 $\pm$ 0.090 & 0.21 $\pm$ 0.02 \\
 & 5.292 & 0.537 $\pm$ 0.007 & 1.93 $\pm$ 0.03 & 2.4 $\pm$ 0.2 & 3.789 $\pm$ 0.005 & 1.0 $\pm$ 0.3 & 0.27 $\pm$ 0.02 \\
 & 6.074 & 0.93 $\pm$ 0.02 & 2.13 $\pm$ 0.03 & 2.9 $\pm$ 0.3 & 6.609 $\pm$ 0.008 & 0.20 $\pm$ 0.05 & 0.26 $\pm$ 0.01 \\
 & 6.460 & 0.54 $\pm$ 0.02 & 1.52 $\pm$ 0.02 & 2.1 $\pm$ 0.4 & 5.91 $\pm$ 0.01 & 0.16 $\pm$ 0.04 & 0.20 $\pm$ 0.01 \\
 & 6.591 & 1.52 $\pm$ 0.03 & 3.11 $\pm$ 0.06 & 4 $\pm$ 1 & 54.9 $\pm$ 0.2 & 0.10 $\pm$ 0.03 & 0.24 $\pm$ 0.02 \\
 & 7.196 & 0.96 $\pm$ 0.02 & 1.92 $\pm$ 0.05 & 2.9 $\pm$ 0.5 & 4.25 $\pm$ 0.03 & 0.04 $\pm$ 0.02 & 0.21 $\pm$ 0.02
\end{tabular}
\end{ruledtabular}
\end{table*}

\section{Microwave Setup}
In Fig.~\ref{fig:wiring_diagram}, we give a schematic of the microwave wiring in our Janis JDry 250 dilution refrigerator. A Keysight PNA N5222B vector network analyzer (VNA) transmits signals down input lines A and B, with 60 dB of discrete attenuation supplied by XMA 2082-6418-dB-CRYO cryogenic attenuators. We estimate an additional 10 dB of attenuation from internal and external line losses.

Two Quinstar QCE-060400CM00 circulators separate input and output paths outside of two six-to-one Radiall 583 microwave switches. Additional directionality on the output lines is achieved with two pairs of 20 dB Quinstar QCI-080090XM00 isolators. Two LNF-LNC4\_8C high electron mobility transistor (HEMT) amplifiers with average noise temperatures of 1.5 K and gain of 40 dB, mounted at the 3 K stage, set the noise floor of our experiments. Room temperature low noise amplifiers (Miteq AFS4-04001200-48-20P-4), provide an additional 30 dB of gain on the receiver side. DC-blocks (Mini-Circuits BLKD-183-S+) decouple DC currents from the input and output lines at room temperature. 
\begin{figure*}[b]
    \centering
    \includesvg[width=120mm,%
    inkscapelatex=false]%
    {microwave_wiring_diagram.svg}
    \caption{Microwave wiring diagram. Passive and active components at each stage of a Janis JDry250 cryogen-free dilution refrigerator used to measure the devices reported in this work. Radiall six-to-one switches, labeled A and B, allow for multiple samples to be measured on one pair of lines during a single cooldown.  Red lines indicate the signal path to measure transmission $S_{21}$ through the devices under test (DUT): LGS1, LGS2, SGS1, and SGS2 resonator chips.}
    \label{fig:wiring_diagram}
\end{figure*}

\section*{References}
\bibliography{TaGrowth2023.bib}